\documentclass[]{aa}  
\usepackage{graphicx}
\usepackage{txfonts}
\begin{document}
%

   \title{Depletion and low gas temperature in the L183 prestellar
   core\,: the N$_2$H$^+$ - N$_2$D$^+$ tool\thanks{Based on
   observations made with the IRAM 30-m and the CSO 10-m. IRAM is
   supported by INSU/CNRS (France), MPG (Germany), and IGN (Spain).}}

   \author{L. Pagani
          \inst{1}
          \and
          A. Bacmann\inst{2}
          \and
          S. Cabrit\inst{1}
          \and
          C. Vastel\inst{3}
          }

   \offprints{L. Pagani}

   \institute{LERMA \& UMR 8112 du CNRS, Observatoire de Paris, 61, Av. de l'Observatoire
   F-75014 Paris\\ \email{laurent.pagani@obspm.fr}
\and Universite Bordeaux 1; CNRS; OASU; UMR 5804, Floirac F-33270
\and Centre d'Etude Spatiale des Rayonnements, 9 avenue du Colonel Roche, 
BP 4346, F-31029 Toulouse Cedex, France}

   \date{Received 31/10/2006; accepted 26/01/2007}

 
  \abstract
  {The study of pre-stellar cores (PSCs) suffers from a lack of undepleted
  species to trace the gas physical properties in their very dense inner parts.}
{We want to carry out detailed modelling of N$_2$H$^+$ and N$_2$D$^+$
cuts across the L183 main core to evaluate the depletion of these species
and their usefulness as a probe of physical conditions in PSCs.}
{We have developed a non-LTE (NLTE) Monte-Carlo code treating the 1D radiative
   transfer of both N$_2$H$^+$ and N$_2$D$^+$, making use of recently published
   collisional coefficients with He between individual hyperfine levels.
   The code includes line overlap between hyperfine transitions. 
   An extensive set of core models is calculated and
   compared with observations. Special attention is paid to the issue
   of source coupling to the antenna beam.}
   {The best fitting models indicate that i) gas in the core center is very cold
   (7$\pm$ 1~K) and thermalized with dust, ii) depletion of N$_2$H$^+$ does occur,
   starting at densities 5--7\,$\times$\,10$^5$ cm$^{-3}$ and
   reaching a factor of 6$^{+13}_{-3}$ in abundance, iii) deuterium
   fractionation reaches $\sim$70\% at the core center, and iv) the
   density profile is proportional to r$^{-1}$ out to $\sim$4000 AU, and
   to r$^{-2}$ beyond.}
   {Our NLTE code could be used to
   (re-)interpret recent and upcoming observations of N$_2$H$^+$ and
   N$_2$D$^+$ in many pre-stellar cores of interest, to obtain better
   temperature and abundance profiles.}


   \maketitle
%

\section{Introduction}
Understanding star formation is critically dependent upon the
characterisation of the initial conditions of gravitational collapse,
which remain poorly known. It is therefore of prime importance to
study the properties of pre-collapse objects, the so-called
pre-stellar cores (hereafter PSCs). As bolometers and infrared
extinction maps have been unveiling PSCs through their dust component,
it has become clear that depletion of molecules onto ice mantles is
taking place inside these cores, preventing their study with the usual
spectroscopic tools.  In most PSCs, very few observable species seem
to survive in the gas phase in the dense and cold inner parts, namely
N$_2$H$^+$, NH$_3$, H$_2$D$^+$ and their isotopologues (e.g. Tafalla
et al. \cite{Tafalla02}). In a few cases, it is advocated that even
N-bearing species also deplete (e.g. B68: Bergin et
al. \cite{Bergin02}; L1544: Walmsley et al. \cite{Walmsley04}; L183:
Pagani et al. \cite{Pagani05}, hereafter PPABC). 

\begin{figure*}
\includegraphics[width=11.cm,angle=-90]{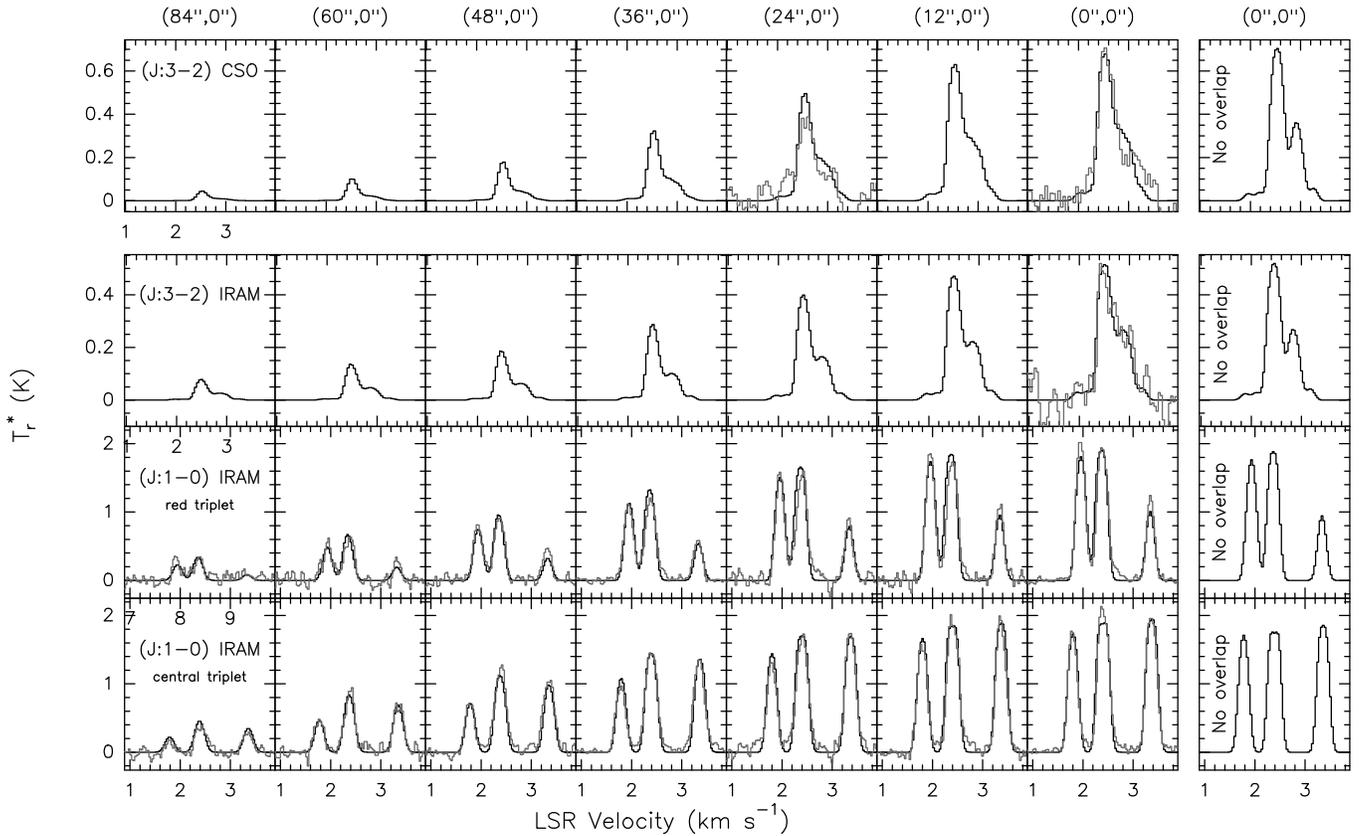}
\caption{N$_2$H$^+$ observations (grey lines) compared to our best
 radiative transfer model (see Fig \ref{ratio}a-b). Each
column corresponds to a spatial offset (indicated above the top box)
from the main dust peak. The top row contains the CSO 10-m data, the
other rows the IRAM 30-m data. The (J:1--0) line is split between the
bottom row (central triplet) and the next row ('red' triplet); the
isolated 'blue' component is not shown. The rightmost column
represents the fit for the central position with a NLTE code without
overlap for the same cloud parameters.}
\label{n2hp}
\end{figure*}

Among the above three species, H$_2$D$^+$ is thus the only one not to
deplete. However, it has only one (ortho) transition observable from
the ground, that moreover requires excellent sky conditions. The para
form ground transition at 1.4 THz should not be detectable in emission
in cores with T$_\mathrm{kin} \leq$ 10 K. Therefore H$_2$D$^+$ is
useful for chemical and dynamical studies, but brings little
information on gas physical conditions. NH$_3$ inversion lines at 23
GHz can provide kinetic temperature measurements as long as higher lying
non-metastable levels are not significantly populated
(Walsmley \& Ungerechts
\cite{Walmsley83}). However, because the critical density of the NH$_3$ (1,1)
inversion line is only $\sim$2000 cm$^{-3}$, 
 this tracer may have substantial contribution from
external, warmer layers,  not representative of
the densest parts of PSCs. The third species, N$_2$H$^+$, appears very
promising\,: it has the strong advantage of having mm transitions with
critical densities in the range 0.5--70\,$\times$\,10$^6$ cm$^{-3}$
and intense hyperfine components for the (J:1--0) line in typical PSC
conditions. The ratio of hyperfine components gives an estimate of the
opacity and excitation temperature of the line. Fitting both the
hyperfine ratios and the (J:1--0) to (J:3--2) ratio with a rigorous
NLTE model should thus bring strong constraints on both
T$_\mathrm{kin}$ and n(H$_2$). Collisional coefficients (with He)
 between individual hyperfine sublevels
have become available recently
(Daniel et al. \cite{Daniel05}). However, current excitation models
(Daniel et al. \cite{Daniel06a}) do not take into account line
overlap, which limits their accuracy.  Another question that remains
is whether N$_2$H$^+$ is indeed able to probe the central core
regions, despite the depletion effects which have been reported.

In this paper, we introduce a new NLTE Monte-Carlo 1D
code\footnote{The Fortran code is available upon request to the
author} treating N$_2$H$^+$ and N$_2$D$^+$ radiative transfer with
line overlap, and apply it to detailed analysis of the main PSC in
L183, a clear-cut case of N$_2$H$^+$ depletion (cf. PPABC).  We
demonstrate the capability of the model to constrain physical
conditions inside the PSC (temperature, density profile, abundance and
depletion, deuterium fractionation). In particular we show that the
gas is very cold at the core center and thermalized
with the dust,  and that N$_2$D$^+$ appears a very useful tracer
of physical conditions in the innermost core regions.

\section{Observations and analysis}
\label{sec:obs}

\begin{table*}
      \caption[]{Spectral parameters of each line observed towards the
      PSC center position ($\alpha$(2000) = 15h54m08.5s,
      $\delta$(2000)= $-$2\degr 52\arcmin 48\arcsec) with
      V$_\mathrm{LSR}$ = 2.3672(2) km\,s$^{-1}$ (from the NH$_3$ (1,1) measurement)\,: noise level, 
reference
      observing frequency, intrinsic linewidth of hyperfine
      transitions and total opacity (both from the CLASS HFS routine), relative velocity and
      intensity of the main hyperfine groups (from gaussian fits).
      Values in parentheses are 1$\sigma$ uncertainties on the last
      digit.  For comparison, the T$_\mathrm{R}^*$ ratios of hyperfine
      groups given by their statistical weights are listed in
italics. The detailed hyperfine structure of the 3 species is
described in Caselli et al. (\cite{Caselli95}) for N$_2$H$^+$, Dore et
al. (\cite{Dore04}) for N$_2$D$^+$ and Kukolich (\cite{Kukolich67}) for NH$_3$
      }
\label{data}
    \begin{math}
        \begin{array}{cccccccccccccccccc}
            \hline
            \noalign{\smallskip}
&\multicolumn{3}{c}{\rm{N_2H^+ (J:1-0)}}&&
&\multicolumn{3}{c}{\rm{N_2H^+(J:3-2)(IRAM)}}&\hspace{0.1cm}&\multicolumn{3}{c}{\rm{N_2H^+(J:3-2)
(CSO)}}&\\
\mathrm{Noise\,(1\sigma)\,:}&\multicolumn{3}{c}{28~ \rm{mK}}&&&\multicolumn{3}{c}{59~ \rm{mK}}&&
\multicolumn{3}{c}{64~ \rm{mK}}&\\
\mathrm{Ref.~frequency\,:}&\multicolumn{3}{c}{93.173764\,\rm{GHz}}&&&\multicolumn{3}{c}{279.511832
\,\rm{GHz}}&&\multicolumn{3}{c}{279.511832\,\rm{GHz}}&\\
\mathrm{Intrinsic~FWHM\,:}&\multicolumn{3}{c}{0.195(1)~ \rm{km\,s^{-1}}}&&&\multicolumn{3}{c}{0.18(2)
~ \rm{km\,s^{-1}}}&&\multicolumn{3}{c}{^\mathrm{a}0.28(4)~ \rm{km\,s^{-1}}}&\\
\mathrm{Total~opacity\,:}&\multicolumn{3}{c}{21.8(5)}&&&\multicolumn{3}{c}{10(2)}&&\multicolumn{3}{c}
{2(1)}&\\
            \noalign{\smallskip}
\cline{2-4}\cline{7-9}\cline{11-13}
            \noalign{\smallskip}
\mathrm{Hyperfine}& V_\mathrm{rel.} & \rm{T_R^*} & Area &\multicolumn{2}{c}{\rm{T^*_R~ Ratio}}  & V_
\mathrm{rel.}  & \rm{T_R^*} & Area &  & V_\mathrm{rel.} & \rm{T_R^*} & Area \\
\mathrm{Group}&\rm{(km\,s^{-1})}&\rm{(K)}&\rm{(K\,km\,s^{-1})}&{\rm obs}&{\rm weights}&\rm{(km\,s^
{-1})}&\rm{(K)}&\rm{(K\,km\,s^{-1})}&&\rm{(km\,s^{-1})}&\rm{(K)}&\rm{(K\,km\,s^{-1})}\\
            \noalign{\smallskip}
            \hline
            \noalign{\smallskip}
1 &-8.009(7)&2.05(3)&0.47(3)&0.932 & {\it 0.428}&-2.58(1)&0.17(6)&0.013(5)&&-2.63(2)&0.20(6)&0.025
(8)\\
2 &-0.611(8)&1.81(3)&0.42(3)&0.822 & {\it 0.428}&0.00(1)&0.50(6)&0.29(2)&&0.00(1)&0.67(6)&0.37(2)\\
3 &0.000(7)&2.20(3)&0.60(3)&1 &{\it 1}&4.70(8)&2.25(6)&0.014(5)&&4.68(3)&2.12(6)&0.028(9)\\
4 &0.956(7)&2.08(3)&0.54(3)&0.945 & {\it 0.714}\\
5 &5.546(7)&2.12(3)&0.47(3)&0.964 & {\it 0.428}\\
6 &5.983(8)&2.04(3)&0.53(3)&0.927& {\it 0.714}\\
7 &6.94(1)&1.26(3)&0.24(3)&0.573 & {\it 0.143}\\ 
            \noalign{\smallskip}
            \hline
            \noalign{\smallskip}
&\multicolumn{3}{c}{\rm{N_2D^+ (J:1-0)}}&&&\multicolumn{3}{c}{\rm{N_2D^+
(J:2-1)^\mathrm{b}}}&&\multicolumn{3}{c}{\rm{N_2D^+
(J:3-2)}}&\\
\mathrm{Noise\,(1\sigma)\,:}&\multicolumn{3}{c}{31~ \rm{mK}}&&&\multicolumn{3}{c}{26~ \rm{mK}}&&
\multicolumn{3}{c}{43~ \rm{mK}}&\\
\mathrm{Ref.~frequency\,:}&\multicolumn{3}{c}{77.109616\,\rm{GHz}}&&&\multicolumn{3}{c}{154.217182
\,\rm{GHz}}&&\multicolumn{3}{c}{231.321917~\rm{GHz}}&\\
\mathrm{Intrinsic~FWHM\,:}&\multicolumn{3}{c}{0.226~ (3)~ \rm{km\,s^{-1}}}&&&\multicolumn{3}{c}
{0.185~ (2)~ \rm{km\,s^{-1}}}&&\multicolumn{3}{c}{0.18~ (2)~ \rm{km\,s^{-1}}}&\\
\mathrm{Total~opacity\,:}&\multicolumn{3}{c}{4.7(3)}&&&\multicolumn{3}{c}{4.9(1)}&&\multicolumn{3}{c}
{1.5(7)}&\\
            \noalign{\smallskip}
\cline{2-4}\cline{7-9}\cline{11-13}
            \noalign{\smallskip}
& V_\mathrm{rel.}  & \rm{T_R^*} & Area &\multicolumn{2}{c}{\rm{T^*_R~ Ratio}}  & V_\mathrm{rel.} & \rm
{T_R^*} & Area &  & V_\mathrm{rel.} & \rm{T_R^*} & Area \\
&\rm{(km\,s^{-1})}&\rm{(K)}&\rm{(K\,km\,s^{-1})}&{\rm obs}&{\rm weights}&\rm{(km\,s^{-1})}&\rm{(K)}&\rm
{(K\,km\,s^{-1})}&&\rm{(km\,s^{-1})}&\rm{(K)}&\rm{(K\,km\,s^{-1})}\\
            \noalign{\smallskip}
            \hline
            \noalign{\smallskip}
1 &-9.697 (8)&0.77(3)&0.15(1)&0.7&{\it 0.428}&-5.35(2)&0.16(3)&0.137(5)&&-3.26(1)&<0.12\\
2 &-0.763 (8)&0.76(3)&0.15(1)&0.691&{\it 0.428}&-0.759 (8)&0.21(3)&0.080 (3)&&0.000(5)&0.72(4)
&0.186(8)\\
3 &0.000 (7)&1.10(3)&0.33(2)&1&{\it 1}&0.000 (1)&1.37(3)&0.513 (3)&&0.46(2)&0.19(4)&0.06(1)\\
4 &1.146 (8)&0.92(3)&0.25(2)&0.836&{\it 0.714}&0.646 (4)&0.40(3)&0.139 (3)&&2.47(3)&0.11(4)&0.028
(7)\\
5 &6.65 (2)&0.55(3)&0.16(2)&0.5&{\it 0.428}&2.5(1)&0.08(3)&0.02(2)&&\\
6 &7.19 (1)&0.90(3)&0.23(2)&0.818&{\it 0.714}&2.97(2)&0.43(3)&0.08(2)&&\\
7 &8.34 (2)&0.34(3)&0.05(2)&0.309&{\it 0.143}&3.57(5)&0.26(3)&0.16(3)&&\\
            \noalign{\smallskip}
            \hline
            \noalign{\smallskip}
&\multicolumn{3}{c}{\rm{NH_3 (1,1)}}&&&\multicolumn{3}{c}{\rm{NH_3 (2,2)}}\\
\mathrm{Noise\,(1\sigma)\,:}&\multicolumn{3}{c}{74~ \rm{mK}}&&&\multicolumn{3}{c}{76~ \rm{mK}}&\\
\mathrm{Ref.~frequency\,:}&\multicolumn{3}{c}{23.6944957~\rm{GHz}}&&&\multicolumn{3}{c}
{23.7226332~\rm{GHz}}\\
\mathrm{Intrinsic~FWHM\,:}&\multicolumn{3}{c}{0.195~ (1)~ \rm{km\,s^{-1}}}&&&\multicolumn{3}{c}{0.20~ 
(1)~ \rm{km\,s^{-1}}}&\\
\mathrm{Total~opacity\,:}&\multicolumn{3}{c}{24.2(4)}&&&\multicolumn{3}{c}{0.1(7)}\\
            \noalign{\smallskip}
\cline{2-4}\cline{7-9}
            \noalign{\smallskip}
& V_\mathrm{rel.} & \rm{T_R^*} & Area &\multicolumn{2}{c}{\rm{T^*_R~ Ratio}}  & V_\mathrm{rel.} & \rm
{T_R^*} & Area \\
&\rm{(km\,s^{-1})}&\rm{(K)}&\rm{(K\,km\,s^{-1})}&{\rm obs}&{\rm weights}&\rm{(km\,s^{-1})}&\rm{(K)}&\rm
{(K\,km\,s^{-1})}\\
            \noalign{\smallskip}
            \hline
            \noalign{\smallskip}
1&-19.504(6)&2.62(7)&0.89(4)&0.922&{\it 0.363}&-0.005(5)&0.73(8)&0.152(7)\\
2& -7.814(5)&2.68(7)&0.69(3)&0.944&{\it 0.273}\\
3& -7.255(8)&1.95(7)&0.57(3)&0.687&{\it 0.182}\\
4& -0.17(1)&2.84(7)&1.15(9)&1&{\it 1}\\
5&  0.30(2)&2.78(7)&1.0(1)&0.979&{\it 0.636}\\
6&  7.465(9)&2.60(7)&0.72(5)&0.915&{\it 0.303}\\
7&  7.89(1)&1.91(7)&0.48(5)&0.673&{\it 0.152}\\
8& 19.318(9)&2.48(7)&0.62(4)&0.873&{\it 0.242}\\
9& 19.85(1)&1.72(7)&0.42(5)&0.606&{\it 0.121}\\
            \noalign{\smallskip}
            \hline
         \end{array}
    \end{math}
\begin{list}{}{}
\item[$^{\mathrm{a}}$] CSO observations are done with a resolution
around 0.1 km\,s$^{-1}$ (or more) which explains this larger value
\item[$^{\mathrm{b}}$] The hyperfine structure is too complex to be detailed
completely. Only the main groups of hyperfine transitions are given
\end{list}
\end{table*}

The main dust core in \object{L183}  (Pagani et
al. \cite{Pagani04}) was observed at the IRAM 30-m telescope in
November 2003, July 2004 and October 2006.  Spectra were taken on a
12\arcsec\ grid in an East-West strip across the core (centered at
$\alpha$(2000) = 15h54m08.5s, $\delta$(2000)= $-$2\degr 52\arcmin
48\arcsec). Symmetric eastern and western spectra were averaged out to
$\pm$ 48\arcsec\ to give a more representative radial profile.
A small anti-symmetric velocity shift of a few tens of m s$^{-1}$
was noticed on either sides of the PSC center at distances beyond
$\pm$ 30\arcsec, possibly indicative of rotation (see
Section~\ref{sec:coremodel}). This shift was compensated for when averaging
eastern and western spectra, in order not to artificially broaden the
lines.  Beyond 48\arcsec, only eastern positions are considered, as
the western side is contaminated by a separate core (Peak 3; Pagani et
al. 2004).  Lines of N$_2$H$^+$ (J:1--0), (J:3--2) and of N$_2$D$^+$
(J:1--0), (J:2--1) and (J:3--2) were observed in Frequency
switching mode, with velocity sampling 30--50 m\,s$^{-1}$ and
T$_\mathrm{sys}$ ranging from 100 K at 3mm up to 1000 K at 1mm.  The
N$_2$H$^+$ (J:2--1) line at 186 GHz was not observed, as it lies only
3 GHz away from the telluric water line at 183 GHz, hence its
usefulness to constrain excitation conditions would be limited by
calibration sensitivity to even tiny sky fluctuations. The problem
does not apply to N$_2$D$^+$ which lies a factor of 1.2 lower in
frequency. Spatial resolution ranges from 33\arcsec\ at 77 GHz to
9\arcsec\ at 279 GHz. Additional CSO 10-m observations of N$_2$H$^+$
(J:3--2) were obtained in June 2004 at selected
positions. Observations were done in Position Switch mode with the
high resolution AOS (48 kHz sampling) and a T$_\mathrm{sys}$ around
600 K. A major interest to observe the (J:3--2) line with the CSO is
that the beam size and efficiency is very similar to the 30-m values
for the (J:1--0) line, thus almost canceling out beam correction
errors in the comparison between the two. It is thus a useful
constraint for radiative transfer modelling, even though the higher
resolution (9\arcsec) 30-m (J:3--2) observations will be essential
to constrain abundances and physical conditions in the innermost core
regions. Standard reduction techniques were applied 
using the CLASS reduction package\footnote{http://www.iram.fr/IRAMFR/GILDAS/}.
 
Complementary observations of NH$_3$ (1,1) and (2,2) inversion lines
towards the PSC center were also obtained at the new Green Bank 100-m
telescope (GBT) in November 2006 with velocity sampling of 20
m\,s$^{-1}$ and a typical T$_\mathrm{sys}$ of 50 K, in Frequency
Switch mode. The angular resolution ($\sim$35\arcsec) is close to that
of the 30-m in the low frequency (J:1-0) N$_2$D$^+$line.
The main beam efficiency of the GBT at 23 GHz is between 0.95 and 1, hence
no correction was applied to put the spectra on T$_\mathrm{R}^*$ scale.

Table \ref{data} summarizes the main observational parameters
of the lines observed towards the core center\,: noise level of the
spectrum (rms), rotational line total opacity and intrinsic FWHM width of individual hyperfine
transitions (as fitted by the CLASS HFS routine, which assumes equal
excitation temperature for all sublevels), and relative velocity centroids and
intensities of the main hyperfine groups in our spectra (derived from
gaussian fits). Note that the hyperfine groups are always broader
than the intrinsic linewidth derived by CLASS, due to optical depth
effects and/or to the presence of adjacent hyperfine components too
close to be spectroscopically separated.

To compare with models, beam efficiency correction is an important
issue. Since L183 (as well as other PSCs) is very extended compared to
the primary beam of 10\arcsec--30\arcsec, the T$_\mathrm{mb}$
intensity scale is inadequate, as it will overcorrect for
source-antenna coupling and thus overestimate the true source surface
brightness\footnote{Teyssier et al.  (\cite{Teyssier02}) provide
appropriate corrections for 30-m data, but only for uniform circular
disk sources, and for data taken before the 1998 surface
readjustment.}  (cf. PPABC). To avoid this problem, observed spectra
are corrected only for moon efficiency ($\sim$ T$_\mathrm{R}^*$
scale), while models are convolved by the full beam pattern of the 30m
telescope (cf. Greve et al. \cite{Greve98})\footnote{The last surface
readjustment of the 30-m in 1998 has not been characterized in detail,
therefore we have scaled down the error beam coupling coefficients
given in Greve et al. (\cite{Greve98}) to retrieve the new, improved
beam efficiencies}, yielding intensities on the same T$_\mathrm{R}^*$
scale. For the CSO, we also correct the data for moon 
efficiency ($\eta_\mathrm{moon}$ $\approx$ 0.8 at 279 GHz) and as the
details of the beam pattern are not known, the model output is
convolved with a simple gaussian beam. Because the CSO main beam coupling
is very good at 1.1 mm ({$\eta_\mathrm{MB} \geq$ 0.7}), the
uncertainty of the correction should be limited. 

In order to treat correctly the beam coupling to the source, we also
take into account the fact that the L183 core is located
within a larger-scale N$_2$H$^+$ filament oriented roughly
north-south, as revealed by previous low resolution maps
(PPABC). This elongation is mostly constant in intensity over $\simeq$
6 arcminutes.  Therefore we approximate the brightness distribution as
a cylinder of length 6\arcmin, with its axis in the plane of the
sky.  The intensity distribution along the east-west section of the
cylinder is taken as the emergent signal along the equator of our
spherical Monte-Carlo model.  We then replicate these values along the
(north-south) cylinder axis before convolving by the antenna beam.
For N$_2$D$^+$, the (unpublished) large-scale map shows a smaller
north-south extent ($\simeq$ 2\arcmin) hence we use a cylinder length
of 2\arcmin\ for convolving our model.

\begin{figure*}
\flushleft
\begin{minipage}[t]{12cm}
\resizebox{12cm}{!}{\includegraphics[width=10.cm,angle=-90]
{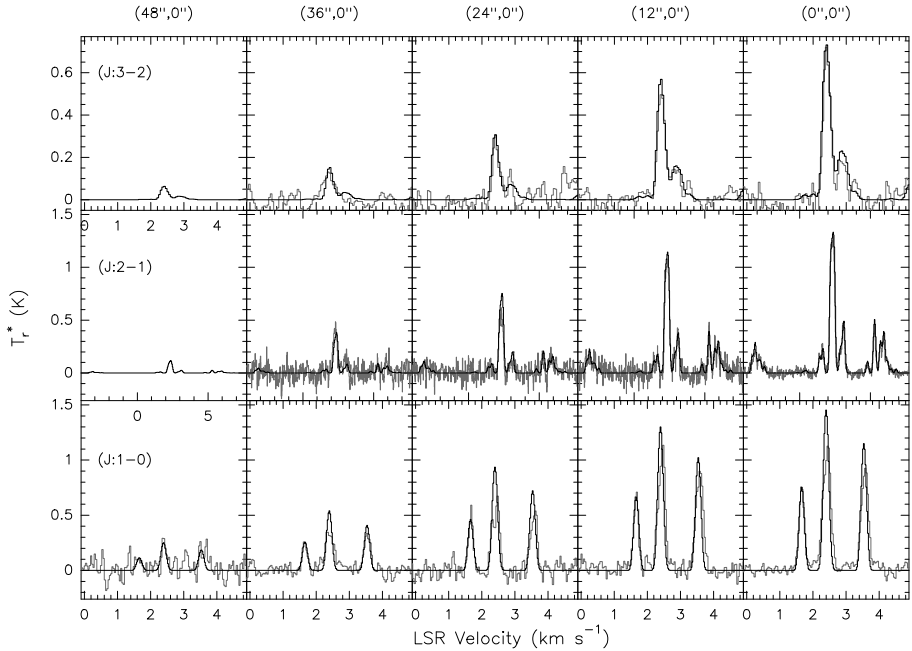}}
\end{minipage}
  \hspace{0.5cm}
  \begin{minipage}[t]{5cm}
  \vspace{5.5cm}
\caption{N$_2$D$^+$ observations (grey lines) compared to our best
radiative transfer model for the largest N$_2$D$^+$ abundance possible
(dotted histogram in Fig \ref{ratio}a). All data are from IRAM 30-m. Only the
central triplet is represented for the (J:1--0) transition}
\label{n2dp}
\end{minipage}
\end{figure*}

In Figs. \ref{n2hp} and \ref{n2dp}, we plot the observations in
T$_\mathrm{R}^*$ compared to our best model (see Section
~\ref{sec:bestfit}), convolved by the antenna beam as explained above.

\section{Modelling and discussion}
\subsection{Radiative transfer code}
\label{sec:montecarlo}
Our spherical Monte-Carlo code is derived from Bernes' code (Bernes
\cite{Bernes}) and has been extensively tested on models of CO
emission from dense cores (e.g. Pagani \cite{Pagani98}). It was recently updated to take
into account overlap between hyperfine transitions occuring at close
or identical frequencies. This is done by treating simultaneously the
photon transfer for all hyperfine transitions of the same rotational
line (see also Appendix in G\'onzalez-Alfonso \& Cernicharo
\cite{Gonzalez93}). Instead of choosing randomly the frequency of the
emitted photon, a binned frequency vector is defined for each rotational transition inside which all
hyperfine transitions are positioned. All bins are filled with their
share of spontaneously emitted photons (or 2.7\,K continuum background
photons) and, during photon propagation, absorption is calculated for
each bin by summing all the hyperfine transition opacities at that
frequency

\begin{displaymath}
\tau(\nu) = \int\kappa(\nu)\mathrm{d}s = \sum_{i}\tau_{i}\phi_{i}(\nu) 
=\int\left(\sum_{i}\kappa_i\phi_{i}(\nu)\right)\mathrm{d}s
\end{displaymath}
where $\kappa_i$ is the absorption coefficient, $\tau_{i}$ the opacity
and $\phi_{i}(\nu)$ the local frequency profile of the i$^{th}$ hyperfine
transition.  The total number of absorbed photons
I$_o(\nu)e^{-\tau(\nu)}$ is then redistributed among hyperfine levels
according to their relative absorption coefficients
\begin{displaymath}
\mathrm{d}I_{i}(\nu) = \frac{\kappa_{i}\phi_{i}(\nu)}{\kappa(\nu)} 
\times I_o(\nu)e^{-\tau(\nu)}
\end{displaymath}

Overlap is treated for all rotational transitions.  Statistical
equilibrium, including collisions, is then solved separately for each
hyperfine energy level.

Collisional coefficients are available for transitions between all
individual hyperfine levels, but were calculated with He as a
collisioner instead of H$_2$ (Daniel et al. \cite{Daniel05}). We
scaled them up by 1.37 to correct for the difference in reduced mass,
but note that this correction is only approximate (see Willey et al.
\cite{Willey02} and Daniel et al. \cite{Daniel06a}).  The code works
for both isotopologues. As one can probably neglect the variation of
the electric dipole moment (Gerin et al. \cite{Gerin01}), the
N$_2$D$^+$ Einstein spontaneous emission coefficients are derived from
those of N$_2$H$^+$ by simply scaling them down by
1.2$^3$. Deexcitation collisional coefficients are kept the same as
for N$_2$H$^+$ (following Daniel et al. \cite{Daniel06b}).

The local line profile in our 1D code takes into account both thermal
and turbulent broadening, as well as any radial and
rotational velocity fields. In principle, rotation introduces a
deviation from spherical symmetry. However, for small rotational
velocities (less than half the linewidth, typically), the excitation conditions do not noticeably
change within a given radial shell, as shown by a previous 2--D
version of this code (Pagani \& Br\'eart de Boisanger,
\cite{Pagani96}). Therefore the (much faster) 1D version remains
sufficiently accurate.

Our Monte-Carlo model shows that excitation temperatures among
individual hyperfine transitions within a given rotational line differ
by up to 15 \%, hence the usual assumption of a single excitation
temperature to estimate the line opacity (e.g. in the CLASS HFS routine)
is not fully accurate as already noticed by Caselli et al. (\cite{Caselli95}). Neglecting line overlap
affects  differentially the excitation temperature of hyperfine components when the opacity is high 
enough, 
mostly decreasing it, but also increasing it in a few cases. For example,
in our best model for the L183 PSC, the excitation
temperatures of individual hyperfine components change by up to 10\%. As shown in the
last column of Fig.~\ref{n2hp}, a noticeable difference between models
with and without overlap appears in the (J:3--2) line shape.

\subsection{Grid of models}
\label{sec:coremodel}

\begin{table*}
      \caption[]{
The 4 main density laws investigated in our models, and the
corresponding temperature and abundance profiles giving the best fit
to N$_2$H$^+$ data in each case. The leftmost model gave the best overall
fit (see Table~\ref{tab:chi2}) and was used to compute the synthetic spectra in
Figs.~1 and 2. Its best N2D+ abundance profile is also listed.
Parameters are held constant inside a given radial shell, except for
the rotational speed which is linearly interpolated.  2.34(6) reads
2.34\,$\times$\,10$^{6}$.  } 
    \begin{math}
        \begin{array}{cccccccccccccccccc}
            \hline
            \noalign{\smallskip}
&&\multicolumn{4}{c}{\rho\propto r^{-1}, r^{-2} 
\,(\rm{Best~model})^{\mathrm{a}}}&&\multicolumn{3}{c}{\rho\propto r^{-1}}&&\multicolumn{3}{c}{\rho\propto r^{-1.5}}&&\multicolumn{3}{c}{\rho\propto r^{-2}}\\
            \noalign{\smallskip}
\cline{3-6}\cline{8-10}\cline{12-14}\cline{16-18}
            \noalign{\smallskip}
\rm{Radius} &  \rm{V.~rot.}&\rm{Density} &T_\mathrm{kin} & 
\rm{Abundance}&\rm{Abundance}& &\rm{Density} &T_\mathrm{kin} & 
\rm{Abundance}&~&\rm{Density} &T_\mathrm{kin} & 
\rm{Abundance}&~& \rm{Density} &T_\mathrm{kin} & 
\rm{Abundance}\\
\rm{A.U.}        &\rm{(km\,s^{-1})} & \rm{(cm^{-3})} &\rm{(K)} & \rm{N_2H^+} & \rm{N_2D^+} & &
\rm{(cm^{-3})} & \rm{(K)} & \rm{N_2H^+} &~& \rm{(cm^{-3})}& \rm{(K)}  & \rm{N_2H^+} &~& \rm{(cm^{-3})}& \rm{(K)}  & \rm{N_2H^+}\\
            \noalign{\smallskip}
            \hline
            \noalign{\smallskip}
  330& 0    & 2.34(6)^{\mathrm{b}}& 7  &2.4(-11) & 2(-11)     & & 2.34(6)^{\mathrm{b}}  & 7 & 7(-11)  & & 3.30(6)^{\mathrm{c}}& 7 & 1(-11)   & & 3.46(6)^{\mathrm{c}}& 9  & 1(-11)  \\
  660& 0    & 2.05(6)^{\mathrm{b}}& 7  &2.4(-11) & 2(-11)     & & 2.05(6)^{\mathrm{b}}  & 7 & 7(-11)  & & 2.89(6)^{\mathrm{c}}& 7 & 1(-11)   & & 3.03(6)^{\mathrm{c}}& 9  & 1(-11)  \\
  990& 0    & 1.55(6)^{\mathrm{b}}& 7  &8.5(-11) & 4(-11)     & & 1.55(6)^{\mathrm{b}}  & 7 & 7(-11)  & & 2.19(6)^{\mathrm{c}}& 7 & 7(-11)   & & 2.29(6)^{\mathrm{c}}& 9  & 3(-11)  \\
 1320& 0    & 1.16(6)		  & 7  &8.5(-11) & 4(-11)     & & 1.16(6) 		& 7 & 7(-11)  & & 1.19(6) 		  & 7 & 7(-11)   & & 1.46(6)& 9  & 3(-11)  \\
 1650& 0    & 9.27(5) 		  & 7  &8.5(-11) & 4(-11)     & & 9.27(5) 		& 7 & 7(-11)  & & 7.74(5) 		  & 7 & 7(-11)   & & 9.09(5)& 9  & 3(-11)  \\
 1980& 0    & 7.73(5) 		  & 7  &8.5(-11) & 4(-11)     & & 7.73(5) 		& 7 & 7(-11)  & & 5.55(5) 		  & 7 & 7(-11)   & & 6.10(5)& 9  & 3(-11)  \\
 2310& 0    & 6.62(5) 		  & 7  &1.1(-10) & 3(-11)     & & 6.62(5) 		& 7 & 1(-10)  & & 4.20(5) 		  & 7 & 4(-10)   & & 4.39(5)& 9  & 1(-10)  \\
 2640& 0    & 5.80(5) 		  & 7  &1.1(-10) & 3(-11)     & & 5.80(5) 		& 7 & 1(-10)  & & 3.34(5) 		  & 7 & 4(-10)   & & 3.54(5)& 9  & 1(-10)  \\
 2970& 0    & 5.15(5) 		  & 7  &1.1(-10) & 3(-11)     & & 5.15(5) 		& 7 & 1(-10)  & & 2.74(5) 		  & 7 & 4(-10)   & & 2.68(5)& 9  & 1(-10)  \\
 3300& 0.01 & 4.64(5) 		  & 7  &1.1(-10) & 3(-11)     & & 4.64(5) 		& 7 & 1(-10)  & & 2.29(5) 		  & 7 & 4(-10)   & & 2.20(5)& 9  & 1(-10)  \\
 3630& 0.01 & 4.21(5) 	 	  & 7  &1.5(-10) & 3(-11)     & & 4.21(5) 		& 7 & 1.5(-10)& & 1.95(5) 		  & 7 & 3(-10)   & & 1.83(5)& 9  & 6(-10)  \\
 3960& 0.01 & 3.54(5) 	 	  & 7  &1.5(-10) & 3(-11)     & & 3.86(5) 		& 7 & 1.5(-10)& & 1.70(5) 		  & 7 & 3(-10)   & & 1.46(5)& 9  & 6(-10)  \\
 4290& 0.02 & 3.01(5)   	  & 7  &1.5(-10) & 3(-11)     & & 3.57(5) 		& 7 & 1.5(-10)& & 1.49(5) 		  & 7 & 3(-10)   & & 1.22(5)& 9  & 6(-10)  \\
 4620& 0.02 & 2.60(5)   	  & 7  &1.5(-10) & 3(-11)     & & 3.31(5) 		& 7 & 1.5(-10)& & 1.32(5) 		  & 7 & 3(-10)   & & 1.07(5)& 9  & 6(-10)  \\
 4950& 0.02 & 2.26(5)   	  & 7  &1.3(-10) & 5(-12)     & & 3.09(5) 		& 7 & 5(-11)  & & 1.18(5) 		  & 7 & 3(-10)   & & 9.27(4)& 9  & 4(-10)  \\
 5280& 0.05 & 1.99(5)   	  & 7  &1.3(-10) & 5(-12)     & & 2.90(5) 		& 7 & 5(-11)  & & 1.07(5) 		  & 7 & 3(-10)   & & 8.5(4) & 9  & 4(-10)  \\
 5610& 0.05 & 1.76(5)   	  & 7  &1.3(-10) & 5(-12)     & & 2.73(5) 		& 7 & 5(-11)  & & 9.67(4) 		  & 7 & 3(-10)   & & 7.20(4)& 9  & 4(-10)  \\
 5940& 0.05 & 1.57(5)   	  & 8  &1.3(-10) & 5(-12)     & & 2.58(5) 		& 8 & 5(-11)  & & 8.82(4) 		  & 7 & 3(-10)   & & 6.34(4)& 9  & 4(-10)  \\
 6667& 0.05 & 1.27(5)   	  & 8  &1(-10)	 & \leq4(-12) & & 2.02(5) 		& 8 & 2.5(-11)& & 8.10(4) 		  & 7 & 1.5(-10) & & 5.25(4)& 9  & 4(-10)\\
 7733& 0.05 & 8.50(4)   	  & 9  &1(-10)	 & \leq4(-12) & & 1.57(5) 		& 9 & 2.5(-11)& & 6.77(4) 		  & 7 & 1.5(-10) & & 4.55(4)& 9  & 4(-10)\\
 8867& 0.05  & 6.50(4)  	  & 10 &1(-10)	 & \leq4(-12) & & 8.50(4)^{\mathrm{d}}  & 10& 2.5(-11)& & 5.42(4) 		  & 7 & 1.5(-10) & & 3.46(4)& 9  & 4(-10)\\
10000& 0.02  & 5.20(4)  	  & 11 &1(-10)	 & \leq4(-12) & & 5.20(4)^{\mathrm{d}}  & 11& 2.5(-11)& & 4.41(4) 		  & 7 & 1.5(-10) & & 2.68(4)& 9  & 4(-10)\\
13333& 0     & 3.45(4)  	  & 12 &1(-10)   & \leq4(-12) & & 3.45(4)^{\mathrm{d}}  & 12& 2.5(-11)& & 3.68(4) 		  & 7 & 1.5(-10) & & 1.70(4)& 9  & 4(-10)\\
            \noalign{\smallskip}
           \hline
         \end{array}
    \end{math}
\begin{list}{}{}
\item[$^{\mathrm{a}}$] $\rho\propto$ r$^{-1}$ (r $<$ 4000~AU), 
$\rho\propto$ r$^{-2}$ (r $>$ 4000~AU)
\item[$^{\mathrm{b}}$] The first three layer densities do not follow a power
law but the density profile derived from the ISOCAM absorption map (Pagani et
al. \cite{Pagani04})
\item[$^{\mathrm{c}}$] Same as note (b) above, but rescaled to maintain
a constant total column density of 1.4$\times$ 10$^{23}$ cm$^{-2}$ (see text).
\item[$^{\mathrm{d}}$] For the pure $\rho\propto$ r$^{-1}$ case,
density is not falling off fast enough to reach ambient cloud
density and we correct for this in the last 3 layers.
\end{list}
\label{tab:model}
\end{table*}

Our core model has an outer radius of 13333 AU $\simeq$ 2\arcmin, fixed
by the maximum width $\simeq$ 4\arcmin of detectable N$_2$H$^+$
emission in the east-west direction across the PSC, as seen in large
scale maps (PPABC). We assume that N$_2$H$^+$ abundance is zero
outside of this radius (due to chemical destruction by CO).

For all models, the microscopic turbulence is set to
$\Delta$v$_\mathrm{turb}$(FWHM) = 0.136\,km\,s$^{-1}$.  This was found
sufficient to reproduce the width of individual hyperfine groups, and
is comparable to the thermal width contribution. A small 
rotational velocity field of a few tens of m\,s$^{-1}$ was further imposed
beyond 3000~AU (see Table~\ref{tab:model}) to reproduce the small
anti-symmetric velocity shift from PSC center observed
at distances beyond $\pm$ 30\arcsec\ (cf. Section~\ref{sec:obs}).
The radial velocity was kept to 0 in all layers.  As
seen in Figs. 1 and 2, this approximation already gives a remarkable fit to
observed line profiles, and is therefore sufficient to derive
the overall temperature and abundance structure. 

We considered a variety of density profiles: single power law profiles
$\rho \propto$ r$^{-1}$, $\rho \propto$ r$^{-1.5}$, $\rho \propto$
r$^{-2}$, as well as broken power laws with $\rho \propto$ r$^{-1}$
out to 4000 AU followed by r$^{-2}$, or by r$^{-3.5}$. For good
accuracy, the density profile was sampled in 330 AU = 3\arcsec\ thick
shells, i.e. 3 to 10 times smaller than our observational beam sizes.
In all cases, divergence at $r = 0$ was avoided by adopting inside r =
990 AU the density slope derived from the
ISOCAM data (Pagani et
al. 2004). The density profiles were then scaled to give the same
total column density N(H$_2$) = 1.4\,$\times$\,10$^{23}$ cm$^{-2}$
towards the PSC center, as estimated from the MAMBO emission map
(Pagani et al. 2004).  The effect of changing the total N(H$_2$) on
the fit results will be discussed in Section~\ref{sec:bestdenstemp}.

We then considered a variety of temperatures spanning the range
6--10 K in the inner core regions. We also tested non-constant temperature laws,
increasing up to 12~K in the outermost layers and/or in the innermost layers.

For each combination of density and temperature laws, a range of
N$_2$H$^+$ abundance profiles was investigated.  To avoid a
prohibitive number of cases, abundances were sampled in 6 concentric
layers only, corresponding to the spatial sampling of our spectra
(except the last, wider layer which encompasses the outermost two
offsets), and multiple maxima/minima were forbidden. This lead to a
total of about 40,000 calculated models with 8 free parameters (6 abundances, 1
temperature profile, 1 density profile).
Table~\ref{tab:model} lists the values of H$_2$ densities for 4
density laws, together with the temperature and abundance profiles
that best fitted the data in each case (cf. next section). 

\subsection{Selection criteria}
\label{sec:selection}

\begin{figure}[th]
\includegraphics[width=5.5cm,angle=-90]{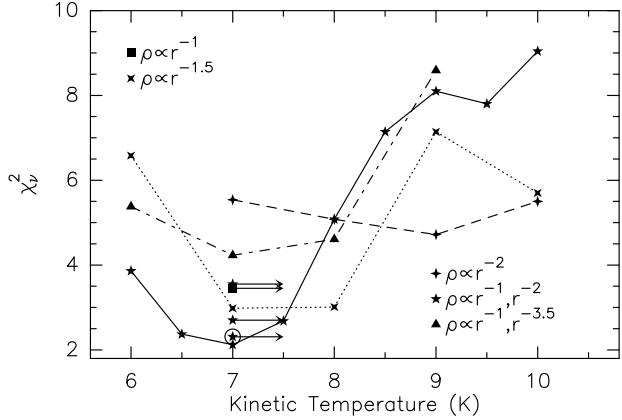}
\caption{Global
$\chi^2_\nu$ for all density profiles and temperature cases. An arrow
indicates that the majority of the model layers is at the given
temperature but that outer layers have rising temperatures. Only one
(circled) solution satisfies all 4 individual $\chi^2$ fitting (see
text) }
\label{fig:chi2}
\end{figure}

Our best fit selection criteria are based on the use of several $\chi^2$ and
(reduced) $\chi^2_{\nu}$ estimates. Because the N$_2$H$^+$ (J:1--0)
data have much more overall weight than the (J:3--2) data (more
isolated hyperfine groups per spectrum, more observed positions,
higher signal-to-noise ratio), computing a single $\chi^2_\nu$ per
model is hardly sensitive to the quality of the (J:3--2) line
fit. However, the (J:1--0) spectra being optically thick, they give
little constraints on the temperature and N$_2$H$^+$ abundance of the
innermost layers (within a 6\arcsec\ radius). Those are only
constrained by the IRAM (J:3--2) observations. The CSO (J:3--2)
spectrum at (24\arcsec,0\arcsec) also adds interesting information
on the density profile. 

To constrain the models we thus
evaluate the goodness of fit independently for the (J:1--0) and
(J:3--2) data, by computing $\chi^2$ values on 4 types of measurements
: 1) full area of each of the 7 (J:1--0) hyperfine components, at each
of the 7 offset positions (49 values). This measurement is sensitive
to the global temperature and abundance profiles. Fitting separately
each hyperfine component allows to discard solutions with overall
identical area but different hyperfine line ratios; 2) area of the 5
central channels (= 0.15 km\,s$^{-1}$) of each hyperfine component for
the same lines (49 values), in order to reject self-absorbed profiles,
i.e. line profiles which have the same total area but are wider and
self-absorbed; 3) total area of the (J:3--2) CSO spectrum at
(24\arcsec,0\arcsec) (1 value). This single measurement helps to
constrain the density profile; 4) total area of the (J:3-2) 30-m
spectrum (1 value) which constrains the abundance (and the temperature
to some extent) at the core center. These four measurements are of
course not completely independent from each other, and trying to
improve one fit by changing a model parameter can degrade one or
several of the others. We used $\chi^2_\nu$ for the first two (with 41
degrees of freedom) and $\chi^2$ for the last two (only one
measurement) so that in all cases $\Delta\chi^2 \approx$ 1 is
equivalent to a 1$\sigma$ deviation. With this choice, a 3$\sigma$
variation is equivalent to $\Delta\chi^2_\nu \approx$ 1.5 for the first
two and $\Delta\chi^2$ = 9 for the last two.


After having run all of the models in our grid, we first used a
single, global $\chi^2_\nu$ combining the four types of measurements
to visualize the global fit quality of the various models. 
Fig. \ref{fig:chi2} plots the global $\chi^2_\nu$ as a function of
temperature, for the various density laws that we investigated. Symbols with arrows
indicate models where temperature rises in the outermost layers. For each 
density-temperature pair, we plot only the smallest $\chi^2_\nu$ obtained 
by adjusting the abundance profile.

It is readily seen that the best fits are found for relatively low
overall core temperatures $\simeq$ 6.5--8~K.  Hence, the (J:1--0) data
alone (which dominate the global $\chi^2_\nu$) set a strong
constraint on the core temperature\,: indeed, the combination of strong
opacity and relatively low intensity in this line requires low
excitation temperature. Because the (J:1--0) line should be
thermalized in the dense center, this rules out kinetic temperatures
above 8~K (for our adopted total column density
of 1.4$\times$ 10$^{23}$ cm$^{-2}$).

To further discriminate among models, we then looked individually at
the four quality indicators described above, in particular those
related to the (J:3-2) lines.  Table~\ref{tab:chi2} lists these 4
values as well as the global $\chi^2_\nu$ for various models selected
from Fig. \ref{fig:chi2}, namely: the best model fits for a given constant core
temperature (from 6 to 10~K), and the best model fits for a given
density law (with corresponding temperature and abundance distributions given in
Table~\ref{tab:model}).

\begin{table*}
      \caption[] {Quality of fit evaluated individually on four different
sets of measurements, and global $\chi_\nu^2$,
for selected models from Fig.~\ref{fig:chi2} (see text).
A difference of 1 corresponds to 1$\sigma$ in all cases.
Values for the best model are in bold face. 
The last rows show the effect of changing only the central N$_2$H$^+$ abundance 
in this best model}
        \begin{tabular}{llcccccc}
            \hline
            \noalign{\smallskip}
&&\multicolumn{2}{c}{(J:1--0)}&&\multicolumn{2}{c}{(J:3--2)}\\
            \noalign{\smallskip}
	\cline{3-4}\cline{6-7}
\noalign{\smallskip}
\multicolumn{2}{c}{model parameters}&  Total area $\chi^2_{\nu}$& 5 channels area
$\chi^2_{\nu}$&& CSO  Total area $\chi^2$&  IRAM  Total area
$\chi^2$&Global $\chi^2_{\nu}$\\
            \noalign{\smallskip}
            \hline
            \noalign{\smallskip}
\multicolumn{8}{c}{Best fit for each temperature (adjusted in abundance)}\\
            \noalign{\smallskip}
            \hline
            \noalign{\smallskip}
T = 6 K &($\rho\propto$ r$^{-1}$,r$^{-2}$)&4.7  &3.8 && 1.5 & 1.0& 3.9 \\
T = 7 K &($\rho\propto$ r$^{-1}$,r$^{-2}$) &1.9 &  2.4 && 12.6 &  4.1& 2.1  \\
T = 8 K &($\rho\propto$ r$^{-1.5}$)&3.1  & 3.2 &&  0.2&   3.4&  2.9\\
T = 9 K &($\rho\propto$ r$^{-2}$)& 4.5 &  5.6&&   7.5&   6.8&  4.7\\
T = 10 K &($\rho\propto$ r$^{-2}$)& 4.8&   6.3&&  16.9& 28.7&  5.5\\
            \noalign{\smallskip}
            \hline
            \noalign{\smallskip}
\multicolumn{8}{c}{Best fit for each density law (adjusted in abundance)}\\
            \noalign{\smallskip}
            \hline
            \noalign{\smallskip}
{\bf $\rho\propto$ r$^{-1}$,r$^{-2}$}&{\bf (T = 7 $\rightarrow$ 12~K)}&{\bf 2.4}&{\bf 2.7}&&{\bf 0.3}&{\bf 0.0}&{\bf 
2.3}\\
$\rho\propto$ r$^{-1}$&(T = 7 $\rightarrow$ 12~K)& 3.7&3.8&&1.2&2.9&3.4\\
$\rho\propto$ r$^{-1.5}$&(T = 7~K)&3.4&3.2&&0.1&0.1&2.9\\
$\rho\propto$ r$^{-2}$&(T = 9~K)&4.5&5.6&& 7.5&6.8&4.7\\
            \noalign{\smallskip}
            \hline
            \noalign{\smallskip}
\multicolumn{8}{c}{Varying the central N$_2$H$^+$ 
abundance at r $<$ 660 AU in the best model}\\
            \noalign{\smallskip}
            \hline
            \noalign{\smallskip}
X(N$_2$H$^+$) = &10$^{-12}$&2.6 &  2.8&&   0.3&   2.0&   2.5 \\
X(N$_2$H$^+$) =&8\,$\times$\,10$^{-12}$&2.6 &  2.8&&   0.3&   1.0&   2.5 \\
{\bf X(N$_2$H$^+$) =}&{\bf 2.4\,$\times$\,10$^{-11}$}&{\bf 2.4}&{\bf 2.7}&&{\bf 0.3}&{\bf 0.0}&{\bf 2.3}\\
X(N$_2$H$^+$) =&8\,$\times$\,10$^{-11}$&2.1 &  2.4 &&  0.5 &  9.2 &2.1\\
            \noalign{\smallskip}
            \hline
            \noalign{\smallskip}
	\end{tabular}
\label{tab:chi2}
\end{table*}

It can be seen in Table \ref{tab:chi2} that for T$_\mathrm{kin}$ = 10~K the
(J:3--2) lines are not well fitted (4 and 5 $\sigma$ deviations for
the CSO and IRAM lines respectively), hence such a high temperature
seems ruled out here. Overall, it is rather difficult to have low
$\chi^2$ values in both the (J:1-0) and the (J:3-2) indicators.  Only
one density-temperature combination (indicated by bold faces in
Table~\ref{tab:chi2}) reaches low values in all 4 individual $\chi^2_\nu$ and
$\chi^2$, each within 1$\sigma$ of their minimum value, among all the
models in Table~\ref{tab:chi2}. This combination is referred to as
our ``best model'' in the following, and its inferred physical
parameters are discussed in the next section. It can also be noticed
in Table \ref{tab:chi2} that the model with T$_\mathrm{kin}$ = 7 K has a slightly
better global $\chi^2_\nu$ than the best model (2.1 compared to 2.3)
but is to be eliminated as the (J:3--2) lines are 2 and 3.5 $\sigma$ off
the mark.

\subsection{Physical conditions in the L183 main core}
\label{sec:bestfit}

\subsubsection{Density and temperature structure}
\label{sec:bestdenstemp}

Our best density-temperature combination (circled 5-branch star symbol
in Fig. \ref{fig:chi2}) has a density law of r$^{-1}$ out to r $\approx$
4000 A.U., consistent with the ISOCAM profile (Pagani et al. 2004) and
a slope of r$^{-2}$ beyond. The temperature is constant at 7~K out to
5600 A.U. and increases up to 12 K at the core edge (see
Table~\ref{tab:model}).  Further increasing the number of warmer layers
failed, as well as introducing warmer layers in the center.  Keeping
the temperature constant at 7~K everywhere significantly worsens the
fit of the (J:3--2) lines (see Table~\ref{tab:chi2}).

The main uncertainty in our temperature determination stems
from the assumed total gas column density through the
core. Decreasing/increasing it by a factor 1.4 (the typical
uncertainty found by Pagani et al. 2004) changes all volume densities
by the same factor. To recover the same N$_2$H$^+$ emission, we find
that the kinetic temperature in our models must be
increased/decreased by $\simeq$1~K (respectively, while the N$_2$H$^+$ abundance
must be scaled accordingly to keep the same column
density). Since the dust temperature in the L183 core is 7.5 $\pm$
0.5~K (Pagani et al. 2004), gas and dust are thermalized inside
this core within the uncertainties.

A different result was found by Bergin et al. (\cite{Bergin06}) in the
B68 PSC, where outer layers emitting in CO are consistent with a
kinetic temperature of 7--8\,K while NH$_3$ measurements indicate a
higher temperature of 10--11\,K in the inner 40\arcsec. This
inward increase in gas temperature was attributed to the lack of
efficient CO cooling in the depleted core center, and requires an
order of magnitude reduction in the gas to dust coupling, possibly due
to grain coagulation (Bergin et al. \cite{Bergin06}).  Such an effect
is not seen in L183, despite strong CO depletion within the inner
1\arcmin = 6600 AU radius. Our best fit temperature law is more consistent
with the thermo-chemical evolution of slowly contracting prestellar
cores with standard gas-dust coupling coefficients, which predicts low
gas temperatures $\simeq$ 6~K near the core center, increasing to
$\simeq$ 14~K near the core surface (Lesaffre et
al. \cite{lesaffre}).


\subsubsection{N$_2$H$^+$ abundance profile}
\label{sec:bestabun}

\begin{figure*}[th]
\includegraphics[width=5.7cm,angle=-90]{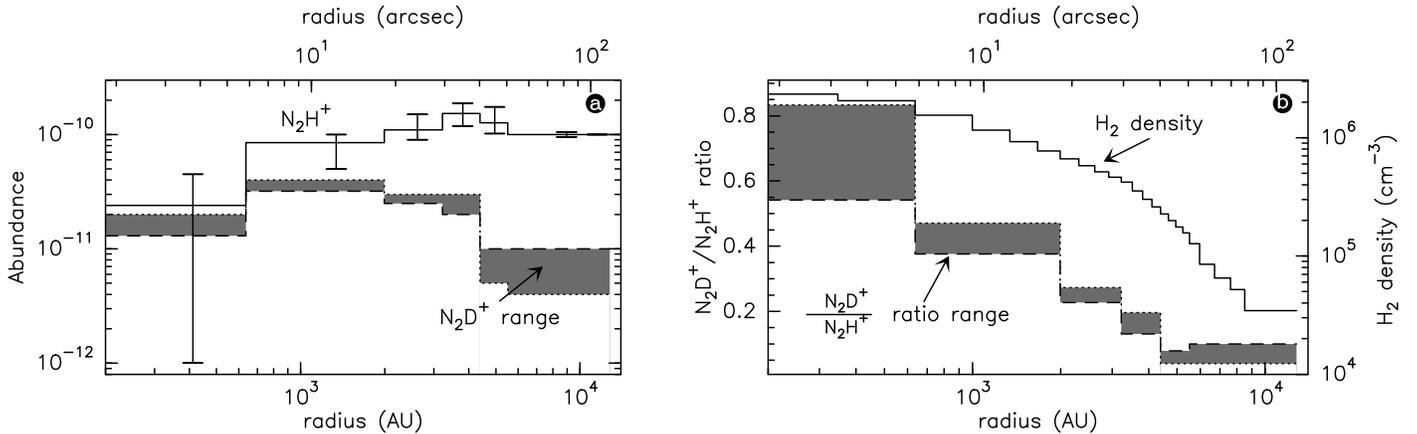}
\caption{{\bf a} N$_2$D$^+$ and N$_2$H$^+$
abundances for the best model. For N$_2$H$^+$, the median abundance
profile is plotted (the corresponding model output is displayed in
Fig. 1). Error bars indicate the range of possible values (see
text). For N$_2$D$^+$, a range of values is given: The upper, dotted
histogram gives the best fit to the (J:2-1) and (J:3-2) lines, and was
used to compute the model in Fig. 2. The dashed histogram gives the
best fit to the (J:1-0) and (J:2-1) lines.  {\bf b} density profile of the best model and
N$_2$D$^+$/N$_2$H$^+$ ratio range }
\label{ratio}
\end{figure*}

With our best density-temperature model, several N$_2$H$^+$ abundance
profiles give equally good fits (within $<$1$\sigma$ on all 4
quality indicators).  From all these acceptable abundance
profiles, we determined a median profile (plotted as a solid histogram
in Fig. \ref{ratio}a), with error bars representing the range of
acceptable values in each layer (though not any combination of these
values does fit the observations).  This median profile, whose values
are listed in  Table~\ref{tab:model}, is used to compute the fit
displayed in Fig. \ref{n2hp}. The total N$_2$H$^+$ column density is
1.2$\pm 0.1$ $\times$ 10$^{13}$ cm$^{-2}$, comparable with the value
reported by Dickens et al. (\cite{Dickens}) and a factor of $\sim$2
below that in Crapsi et al. (\cite{Crapsi05}).

The maximum N$_2$H$^+$ abundance is 1.5$_{-0.3}^{+0.4}$ $\times$
10$^{-10}$, the same as found by Tafalla et al. (\cite{Tafalla04}) in
L1498 and L1517. However, there is a definite drop
in abundance in the inner layers. This drop is imposed by the fit to
the (J:3--2) IRAM central spectrum (9\arcsec\ beam), which is quite
sensitive to variations in the central abundance of N$_2$H$^+$, as
illustrated in the last rows of Table~\ref{tab:chi2}.  In contrast,
the (J:1--0) line is optically thick at the core center and the
$\chi_\nu^2$ is thus not sensitive to this parameter
(cf. Table~\ref{tab:chi2}).

Exploring a broad range of abundance profiles, we find only a limited
range of possible N$_2$H$^+$ abundances in the inner layer of
6\arcsec\ radius (r $<$ 660~AU): values above
4.5\,$\times$\,10$^{-11}$ always produce too strong (J:3--2) emission
in the central IRAM beam ($\chi^2 > 1$), while abundances below
10$^{-12}$ give a (J:3--2) line marginally too weak. We thus derive 
a range of 2.4 $^{+2.1}_{-2.3}$\,$\times$\,10$^{-11}$ for the
central N$_2$H$^+$ abundance. However, we note that our results for N$_2$D$^+$
would favor a value of at least 10$^{-11}$ (see next section).

Compared with the maximum abundance, this gives a volume depletion
factor of $\simeq$ 6$^{+13}_{-3}$ at the core center. The
median abundance drop is less than (but marginally consistent with)
that expected from simple geometrical arguments in PPABC, but it
confirms that the leveling of N$_2$H$^+$ intensity seen across the
dust peak is not due to pure opacity effects. As can be seen from
Fig. \ref{ratio}a-b, depletion starts at a density of
5--7\,$\times$\,10$^5$ cm$^{-3}$ and increases as density goes up, in
agreement with the conclusions of PPABC. The abundance also drops
slightly in the outermost regions, possibly due to partial destruction
by CO.  Indeed, the outermost layers of the model have densities of a
few 10$^4$ cm$^{-3}$ which is the limit above which CO starts to
deplete in this source (PPABC).

\subsubsection{N$_2$D$^+$ abundance profile}
\label{sec:bestn2dp}

As seen in Fig.\ref{n2dp}, good fits to the N$_2$D$^+$ data may be
obtained with the same density and temperature profile as our best
model for N$_2$H$^+$, although it is difficult to reproduce
simultaneously the intensities of both (J:1-0) and (J:3-2) lines. We
plot in Fig. \ref{ratio}a the N$_2$D$^+$ abundance profiles giving the best fit
either for (J:1-0) and (J:2-1) (dashed histogram) or for (J:2-1) and
(J:3-2) (dotted histogram). The latter was used to produce
Fig. \ref{n2dp}. A better fit could be obtained with a temperature of
8\,K instead of 7\,K. This might indicate that the collisional
coefficients with He are somewhat inaccurate to represent those with
H$_2$. Indeed, it has been shown for NH$_3$ that collisional
coefficients with He could differ by a factor up to 4 with respect to
those computed with para-H$_2$ (Willey et al. \cite{Willey02}). A
similar problem may occur here (see also the discussion in Daniel et
al. \cite{Daniel06a}). Therefore we consider that the temperatures are
compatible within the uncertainties on the collisional coefficients.

The N$_2$D$^+$ abundance profile is quite different from that of
N$_2$H$^+$ (cf. Fig. \ref{ratio}a). Its abundance is essentially an
upper limit beyond 6000 A.U.  It increases sharply by about an order
of magnitude in the region between 600 and 4000 A.U., then slightly
drops by a factor of 2--2.5 in the core center, reaching an abundance
between 1.3 and 2\,$\times$\,10$^{-11}$. The low optical depth of the
line ($\tau$ = 0.84 for the strongest hyperfine component,
J$_{FF\arcmin}$:1$_{23}$--0$_{12}$) allows to measure the contribution
of all layers to the emission and to determine with relatively little
uncertainty the abundance profile. In particular, for the chosen
density and temperature profiles, we find that the abundance of N$_2$D$^+$ in the
center of the cloud cannot be below 10$^{-11}$. Hence, we can set
tighter constraints on the N$_2$D$^+$ abundance at the core center
than was possible for N$_2$H$^+$.

The N$_2$D$^+$/N$_2$H$^+$ ratio obtained by comparing the
N$_2$D$^+$ range of abundances to the median N$_2$H$^+$ abundance
profile is plotted in Fig. \ref{ratio}b.  The deuteration ratio varies
from an upper limit $\leq$0.05--0.1 away from the core to a very high
factor of 0.7$\pm$0.12 in the depletion region. This is larger than
what Tine et al. (\cite{Tine}) reported 48\arcsec\ further north in
the same source, and 3--4 times larger than what Crapsi et
al. (\cite{Crapsi05}) report towards the PSC. Deuterium enrichment is
thus very strong and can certainly be linked to the strong and
extended H$_2$D$^+$ line detected towards this source (Vastel et
al. \cite{Vastel06}). If we consider the full range of possible values
for N$_2$H$^+$ itself, the deuteration ratio in the inner layer ranges
from $\sim$0.3 to $\geq$ 20. However, as D$_2$H$^+$ has not been
detected in this source despite its strong H$_2$D$^+$ line (Vastel et
al. \cite{Vastel06}), an enrichment above 1 seems unprobable,
suggesting that the central abundance of N$_2$H$^+$ is probably at
least 10$^{-11}$.

\subsection{Comparison with NH$_3$ temperature estimates}

\begin{figure}[t]
\includegraphics[width=6.2cm,angle=-90]{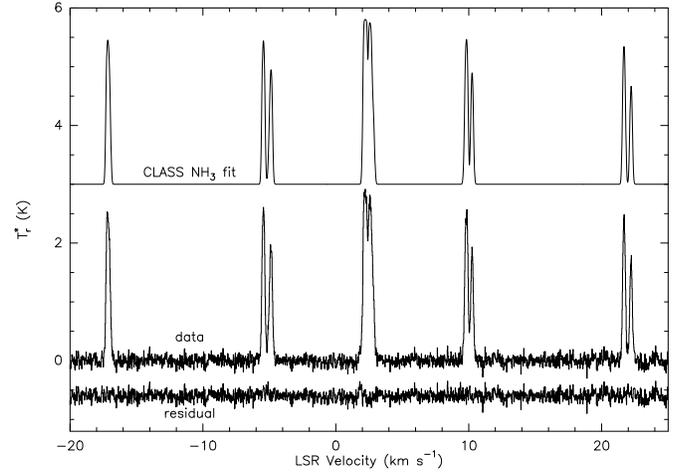}
\caption{GBT NH$_3$ (1,1) inversion line towards the reference position
with CLASS NH3 fit (shifted by +3 K for clarity) and fit residual
(shifted by -0.6 K). The fit yields a total opacity of 24.2 ($\pm$0.4),
and a global velocity of 2.3672 ($\pm$0.0002) km\,s$^{-1}$}
\label{nh311}
\end{figure}

\begin{figure}[ht]
\includegraphics[width=6.2cm,angle=-90]{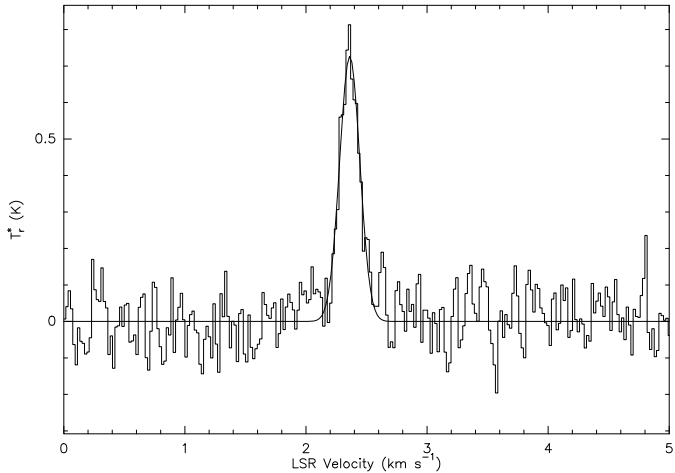}
\caption{GBT NH$_3$ (2,2) inversion line towards the reference position. A
gaussian fit to the main line indicates a velocity of 2.362
($\pm$0.005) km\,s$^{-1}$}
\label{nh322}
\end{figure}

The low kinetic temperatures $\le$8~K inferred from our
N$_2$H$^+$ and N$_2$D$^+$ modelling are somewhat lower than previous
temperature estimates in L183 from NH$_3$ inversion lines, which gave
values in the range 9--10~K (Ungerechts et al. \cite{Ungerechts80}, Dickens et
al. \cite{Dickens}) up to 12~K (Swade et al. \cite{Swade89}). However,
these NH$_3$ spectra were not obtained towards the PSC center itself,
and had relatively low angular resolution for the last two.  We thus briefly reconsider
this issue using our NH$_3$ spectra obtained at the PSC center
position, which also benefit from the higher resolution and beam
coupling of the new GBT. The NH$_3$ spectra were analyzed in the
standard way, as discussed by Ho \& Townes (\cite{Ho83}) and Walmsley
\& Ungerechts (\cite{Walmsley83}).  Using the CLASS NH3 fitting
procedure, we found a total opacity of 24 for the (1,1) inversion
line, and an excitation temperature of 5.5 K assuming a beam
filling-factor of 1 (Figs.~\ref{nh311} \& \ref{nh322}). Making the
usual assumption of constant temperature on the line of sight and of
negligible population in the non-metastable levels, the
intensity ratio of the (2,2) to (1,1) main lines indicates a
rotation temperature of 8.4\,$\pm$0.3\,K which should correspond to a
kinetic temperature of 8.6\,$\pm$0.3\,K.  Hence, NH$_3$ emission
towards the PSC center indicates an only slightly higher kinetic
temperature than N$_2$H$^+$ (8.6\,$\pm$0.3\,K instead of
7\,$\pm$1\,K), almost equal to that obtained with N$_2$D$^+$
($\sim$8\,K).  The discrepancy between NH$_3$ and N$_2$H$^+$ is thus
not as large as originally thought. Both tracers point to very cold
gas in the core,  close to thermal equilibrium with the dust.


Reasons for a possible  difference between NH$_3$ and
N$_2$H$^+$ temperature determinations in L183 include the following\,:
1) the higher sensitivity of NH$_3$ to the warmer outer layers of the
core, since NH$_3$ inversion lines are much easier to thermalize
(n$_\mathrm{crit} \approx$ 2000 cm$^{-3}$) than N$_2$H$^+$ lines.  2)
a slight overestimate in the total column density towards the PSC
(reducing it to 10$^{23}$\,cm$^{-2}$, the temperature has to be raised
to 8 K to compensate for the density decrease and recover the same
N$_2$H$^+$ emission). 3) systematic errors introduced by the use of
collisional coefficients with He instead of H$_2$ for N$_2$H$^+$ (see
Sect. \ref{sec:bestn2dp}). 4)
concerning NH$_3$, standard hypotheses leads to a puzzling
discrepancy between T$_\mathrm{ex}$ = 5.5 K and T$_\mathrm{rot}$ =
8.4~K towards the L183 PSC center, where NH$_3$ inversion lines
should be fully thermalized. Beam dilution has been invoked for giant
molecular clouds (thus increasing T$_\mathrm{ex}$), but it seems
unreasonable to extend this to dense cloud cores (Swade
\cite{Swade89}, and references therein). A Monte-Carlo code (or
equivalent) would be needed to model NH$_3$ taking into account the
strong density gradients present in PSCs, and the possible population
of non-metastable levels at very high densities.

\section{Conclusions}

   \begin{enumerate}
	\item We have presented a new Monte-Carlo code (available upon
request to the author) to compute more realistically the NLTE emission of
N$_2$H$^+$ and N$_2$D$^+$, taking into account both line overlap and
hyperfine structure. This code may be used to infer valuable
information on physical conditions in PSCs.

	\item The best kinetic temperature to explain N$_2$H$^+$
      observations of the L183 main core is 7$\pm$1~K (and $\sim$ 8\,K
      for N$_2$D$^+$) inside 5600~AU, therefore gas appears
      thermalized with dust in this source.

    \item There is no major discrepancy with NH$_3$ measurements which
    also indicate very cold gas (8.6\,$\pm$0.3\,K) towards the PSC.

      \item We have found a noticeable depletion of N$_2$H$^+$ by a
      factor of 6$^{+13}_{-3}$, and of N$_2$D$^+$ by a smaller
      factor of 2--2.5. This smaller depletion is probably due to a
      strong (0.7$\pm$0.12) deuterium fractionation, consistent with
      the detection of H$_2$D$^+$ in this core.

      \item N$_2$D$^+$ should be a useful probe of the innermost core
      regions, thanks to its low optical depth combined with its strong
      enhancement.

   \end{enumerate}

\begin{acknowledgements}
We thank the IRAM direction and staff for their support, S. L\'eon for
his dedicated assistance during pool observing, and F. Daniel for
providing routines to compute the frequencies and A$_{ul}$
coefficients of N$_2$H$^+$. We also thank an anonymous referee and
C.M. Walmsley for suggestions which helped to improve this paper.
\end{acknowledgements}

\end{document}